\documentclass[prd,aps,twocolumn,nofootinbib,longbibliography]{revtex4}
\usepackage{graphicx}
\usepackage{hyperref}
\usepackage{amsmath,amssymb}
\usepackage{xcolor}

\definecolor{darkBlue}{rgb}{0, 0, 0.8}

\hypersetup{
  bookmarksopen=true,     
  colorlinks=true,        
  linkcolor=darkBlue,     
  citecolor=darkBlue,      
  filecolor=darkBlue,      
  urlcolor=darkBlue        
}

\def\be{\begin{equation}}
\def\ee{\end{equation}}
\def\ba{\begin{align}}
\def\ea{\end{align}}

\def\lsim{\raise0.3ex\hbox{$\;<$\kern-0.75em\raise-1.1ex\hbox{$\sim\;$}}}
\def\gsim{\raise0.3ex\hbox{$\;>$\kern-0.75em\raise-1.1ex\hbox{$\sim\;$}}}

\def\theta{\vartheta}

\def\d{{\rm d}}

\def\R{{\cal R}}
\def\V{{\cal V}}

\newcommand\vev[1]{{\left\langle{#1}\right\rangle}}

\newcommand{\aap}{{Astron.\ Astrophys. }}

\newcommand{\mnras}{{Mon.\ Not.\ Roy.\ Astron.\ Soc. }}

\renewcommand{\vec}[1]{\boldsymbol{#1}}

\begin{document}

\title{Hot spots in the neutrino flux created by cosmic rays from Cygnus and Vela?}

\author{M.~Bouyahiaoui$^{1}$\footnote{Now at Max-Planck-Institut f\"ur Kernphysik, Heidelberg, Germany}}
\author{M.~Kachelrie\ss$^{2}$}
\author{D.~V.~Semikoz$^{1,3,4}$}

\affiliation{$^{1}$APC, Universit\'e Paris Diderot, CNRS/IN2P3, CEA/IRFU,
Observatoire de Paris, Sorbonne Paris Cit\'e, 119 75205 Paris, France}
\affiliation{$^{2}$Institutt for fysikk, NTNU, Trondheim, Norway}
\affiliation{$^{3}$INR RAS, 60th October Anniversary prospect 7a, Moscow, Russia }
\affiliation{$^{4}$National Research Nuclear University MEPHI, Kashirskoe highway 31, 115409 Moscow, Russia}

\begin{abstract}
An analysis of 7.5\,years of data in the high-energy starting event sample
has been recently published by the IceCube collaboration. The hottest spot
in a search for neutrino sources was found far above the Galactic plane and
is thus, at first sight, difficult to reconcile with a Galactic origin.
In this work, we calculate the cosmic ray (CR) density around nearby, young
supernova remnants assuming anisotropic diffusion. Combining the obtained
CR densities with the matter distribution deduced from extinction maps, we
find two prominent hot spots: The one close to the most significant
point in the IceCube search for point sources is created by CRs from the
Cygnus loop and has an intensity corresponding to two to four neutrino events.
Another, more extended one may be caused by CRs from Vela if CR
trajectories are sufficiently disturbed by the magnetic field in the
shell around the superbubble Loop~I.  
\end{abstract}

\maketitle

\section{Introduction}

Neutrinos are a unique tool to study dense environments and the non-thermal
universe~\cite{Raffelt:1999tx,Gaisser:1994yf,Kachelriess:2019oqu}.
High-energy neutrinos may be produced together with photons
in hadronic interactions of cosmic rays (CR) close to their sources or during
propagation. Since they travel undisturbed, being neither absorbed as
high-energy photons nor deflected in magnetic fields as charged particles,
they are an ideal tracer of sites where the product of CR and target
density is high. Such places can be either the CR sources themselves or
dense gas clouds close to the sources. Therefore high-energy neutrino
observatories
have the potential to identify the yet unknown sources of CRs---and
this capability has been one of the key motivation for their construction.

The currently largest of these high-energy neutrino telescopes, the IceCube
observatory, has succeeded to establish the existence of a surprisingly large
flux of extraterrestrial high-energy neutrinos~\cite{Aartsen:2016xlq}.
In Ref.~\cite{Abbasi:2020jmh},  the IceCube collaboration has reanalyzed
7.5\,years of data using neutrinos which
interaction vertices are contained inside the fiducial volume of the detector.
The energy spectrum of these ``HESE'' neutrinos is compatible with an unbroken
power law ${\rm d}N/{\rm d}E\propto E^{-\alpha}$ with a spectral index of
$\alpha=2.87\pm ^{0.20}_{0.19}$. Such a steep spectrum challenges most
extragalactic source models. In particular, it requires the production
  of neutrinos in hidden extragalactic sources, since otherwise
the accompanying photons would overshoot the bounds on the diffuse background
of extragalactic gamma-rays~\cite{Berezinsky:2010xa,Murase:2013rfa}.

A guaranteed contribution to the astrophysical neutrino flux are secondary
neutrinos produced in interactions of Galactic ``sea'' CRs with
gas~\cite{Berezinsky:1975zz,Berezinsky:1992wr}.
However, the magnitude of this flux is, even using optimistic parameters,
well below the one observed~\cite{Kachelriess:2014oma,Gaggero:2015xza}.
Moreover, the arrival directions of the astrophysical neutrinos show not
the correlation with the Galactic plane  expected in this scenario. Therefore,
several alternatives which predict a close to isotropic Galactic neutrino
flux were suggested:
Such neutrinos may originate from a large Galactic halo, formed either
by CRs~\cite{Taylor:2014hya,Blasi:2019obb} or heavy dark matter
particles~\cite{Berezinsky:1997hy,Feldstein:2013kka,Esmaili:2013gha}, or
from the Fermi bubbles~\cite{Crocker:2010dg,Lunardini:2011br}. 
Alternatively, a significant contribution to the Galactic neutrino flux may
be rather local, produced by CR interactions at the boundary the Local
Bubble (LB)~\cite{Andersen:2017yyg,Bouyahiaoui:2020rkf}.

Studies of specific Galactic neutrino sources have mostly been based on the
gamma-ray--neutrino connection: Since the production spectra of gamma-rays
and neutrinos in hadronic interactions are correlated, estimates and strict
upper bounds on the neutrino flux can be deduced from gamma-ray observations
for specific CR sources, see e.g.\ Ref.~\cite{Villante:2008qg}.
In the case of young CR sources, the finite, energy-dependent distance
CRs diffuse may lead, however, to a low-energy cutoff in the secondary
spectrum. Therefore gamma-ray observations which are typically limited to
energies ${\cal O}(\rm TeV)$ restrict only weakly the
neutrino flux from such sources at ${\cal O}(100\,\rm TeV)$.  As result,
the brightest spots in the Galactic neutrino sky may be gas clouds
immersed into the CR overdensities close---but not close enough to be visible
in TeV gamma-rays---to nearby young CR sources~\cite{Kachelriess:2014oma}.

In this work, which is an extension and update of
  Ref.~\cite{Bouyahiaoui:2021rhm},
we study if the nearest Galactic CR accelerators may be
visible as neutrino sources. We calculate the CR density around nearby,
young supernova remnants (SNR), assuming anisotropic diffusion as suggested
in Refs.~\cite{Giacinti:2014xya,Giacinti:2015hva,Giacinti:2017dgt}.  The
faster diffusion of CRs along magnetic field lines implies that gas clouds
can give an important contribution to the secondary neutrino flux at larger
distances than in the case of isotropic diffusion, if their perpendicular 
distance to the magnetic field line through the CR source is small. In
addition, the slow diffusion perpendicular to the magnetic field enhances
the CR density close to magnetic field lines connected to the CR sources.
We extract the local matter distribution acting as target for neutrino
production from extinction maps derived in Refs.~\cite{Leike_2019,Leike_2020}.
Combining then this matter map with the calculated CR densities, we find
two prominent hot spots: The one close to the most significant
point in the IceCube search for point sources is created by CRs from the
Cygnus Loop supernova remnant (SNR).
We estimate the neutrino intensity of this hot spot and find it
compatible with two to four events observed by IceCube with energy above
60\,TeV.
The other one, produced by CRs from the Vela SNR, is at first sight
absent in the IceCube HESE sample. However, the CR flux from Vela
should be spread out by the enhanced magnetic field in the wall around the
superbubble Loop~I. We point out that the region around the intersection
of Loop~I with the peak of the  predicted CR intensity has an excess
of observed neutrino events, supporting this interpretation.

\section{Local cosmic ray sources}
\label{source}

We choose to use as potential CR sources SNRs  which are younger than
30\,kyr and
located closer than 1\,kpc from the Sun. Table~\ref{table:tab_sources}
summarizes the available information\footnote{Data extracted from
  \url{http://snrcat.physics.umanitoba.ca/SNRtable.php}.} on these sources.
For the calculation of the CR density around these SNRs, we use the
same approach as in  Refs.~\cite{Bouyahiaoui:2018lew,Bouyahiaoui:2020rkf}:
We calculate the trajectory of individual CRs using the Lorentz equation,
and find the CR density from the time spent by CRs in space  cells.
 We approximate the
Galactic magnetic field by the  Jansson-Farrar
model~\cite{Jansson:2012rt,Jansson:2012pc}, with a reduced turbulent
field as suggested in
Refs.~\cite{Giacinti:2014xya,Giacinti:2015hva,Giacinti:2017dgt}.
Turbulent magnetic fields are generated using nested
grids  as descibed in ~\cite{Giacinti:2011ww,Giacinti:2011uj} which allows
us to include a large dynamic range of spatial scales contributing to the
turbulence.The turbulent magnetic field is taken to be randomly directed with
modes distributed between
$L_{\rm min} = 1$\,AU and $L_{ \rm max} = 25$\,pc according to an isotropic
Kolmogorov power spectrum.

Moreover, we modify this model as described in Ref.~\cite{Bouyahiaoui:2020rkf}
to take into account the Local Bubble.
We apply an exponential damping of the magnetic field inside the
bubble as function of the distance $z$ to the Galactic plane with height
scale $z_{\rm b}= 100 $\,pc. 
We assume that the strength of the regular magnetic field inside the bubble
depends only on the radius $r$ and $z$, setting $B_{\rm in}=0.1 \mu G$ inside
the bubble and $B_{\rm sh}=8 - 12\mu G$ in the wall. Inside the bubble and
the wall, we use a clockwise oriented magnetic field for $y>0$ and an
anticlockwise one for $y<0$. 
We interpolate the transition between different magnetic field regimes by
logistic functions $T(r)$. The width of the two transitions is parameterised 
by $w_{i={1,2 }} $, while $w$ denotes the extension of the wall. We set
\begin{align}
T_{1/2} & = \left[ 1+ \exp\left(-11\,\frac{r-R \pm w/2}{w_{1/2}}\right) \right]^{-1} 
\end{align}
The Sun is assumed to be at the centre of the LB.
In this configuration the turbulent magnetic field strength is set as follows.
For $r > R-w/2$:  
$B_{\rm turb}  =  B_{\rm reg}/2$, and for $r \geq R-w/2$:
$B_{\rm turb}  =  5B_{\rm reg}$, where  we use as numerical values
$w_1 = 1$\,pc, $w_2 = 0.1$\,pc, $w = 2$\,pc and $R = 100$\,pc.

\begin{table}[h!]
\begin{center}
\begin{tabular}{c|c|c|c|}
 Source & Name  & $\tau$/kyr &$d$/kpc \\ 
  \hline
   G065.3+05.7 & -- & 20 & 0.8   \\ 
    \hline
  G074.0-08.5 & Cygnus Loop & 15 & 0.75   \\ 
    \hline
    G106.3+02.7 & Boomrang  & 10 & 0.8    \\ 
    \hline
    G114.3+00.3 & --  & 7.7   & 0.7  \\ 
  \hline
   G160.9+02.6 & HB9 & 5.5  & 0.8   \\ 
  \hline
   G263.9-03.3 & Vela & 11   & 0.29  \\ 
  \hline
   G266.2-01.2 & Vela Jr  & 3.8 & 0.75   \\ 
  \hline
   G330.0+15.0 & Lupus Loop& 23   & 0.33   \\ 
  \hline
   G347.3-00.5 & -- & 1.6   & 1  \\ 
  \hline
   B1737-30 & -- & 20.6  & 0.4   \\ 
\end{tabular}
\end{center}
\caption{Cosmic ray sources considered with average values for their age $\tau$ and distance $d$ from the Sun.  \label{table:tab_sources}}
\end{table}

No detailed model for the time-dependent
escape spectrum of accelerated CRs exist. Since we are interested only in
CRs with the highest energies which are accelerated before the end of 
the Taylor-Sedov phase, we can however assume that these CRs escaped at a
time much smaller than the age $\tau$ of the source. Moreover, we employ the
same injection spectrum for all sources for which we choose motivated by the
results of Ref.~\cite{Bouyahiaoui:2020rkf} a broken power
law in rigidity $\R=E/(Ze)$ with a break at $\R_{\rm br}=2\times 10^{15}$\,V
and an exponential cut-off at $\R_{\rm max}=8 \times 10^{15}$\,V,
\be \label{spec}
\frac{dN_i}{d\R} =  \left\{
    \begin{array}{ll}
      N_i  \,\R^{-2.2} , & \mbox{\quad if \quad} \R < \R_{\rm br} \\
      \tilde N_i  \,\R^{-3.1} \exp(-\R/\R_{\rm max}), & \mbox{\quad if \quad} \R \ge \R_{\rm br}.
    \end{array}
    \right.
\ee
We inject 5.000~protons per energy at the position of each source and
propagate them for the time $\tau$ in our magnetic field model.
Then we calculate the CR density $n(E)$ in each cell of size (6\,pc)$^3$.
The energy injected in CRs by each source is chosen as 
$3.5\times 10^{50}$\,erg in protons and $2.6\times 10^{50}$\,erg in 
helium, respectively, which is compatible with the expectation that 
$\simeq 50$--60\% of the explosion energy are transferred into relativistic
particles in the case of efficient CR 
accelerators~\cite{2001MNRAS.321..433B,Ellison:2003fc,2004MNRAS.353..550B}.

\section{Local matter density from dust maps}
\label{dust}

While dust contributes only a small mass fraction to the interstellar medium
(ISM), it efficiently absorbs and scatters photons with wavelengths in
the visible and ultraviolet range. Therefore the distribution of dust can
be used as a tracer of, e.g., gas  which serves CRs as target for
secondary neutrino and gamma-ray production.

Most efforts to build 3D maps of Galactic dust have been concentrated on
charting dust on large scales. For instance, Ref.~\cite{2019ApJ...887...93G}
mapped three quarters of the sky, 
while Ref.~\cite{Lallement_2018} constructed a map extending out to 3\,kpc
with 25\,pc resolution. These maps
succeeded modeling Galactic dust on large scales, but they failed to recover
features on small scales because of their missing resolution. 
In contrast, the authors of Refs.~\cite{Leike_2019,Leike_2020} concentrated
their study on
the closer neighborhood of the Sun, building a dust map for a
$740^2 \times 540 $\,pc$^3$ cube with a superior resolution of 1\,pc.
Using Gaia, 2MASS, PANSTARRS, and ALLWISE data,
they deduced the G band extinction of five million stars with known
parallaxes. Their results for the 3D distribution of dust are publicly
available as a grid containing the e-folds of extinction per cell.

The extinction due to dust is proportional to the hydrogen column density
along the grid cell as~\cite{2016ApJ...826...66F}
\be
 N_{\rm H} = 2.87 \times 10^{21}\,{\rm cm}^{-2} \;  A_{\rm V}/{\rm mag} . 
\ee
Using moreover $A_{\rm G}/A_{\rm V}=0.789$ as the selective extinction in the
GAIA G~band from Ref.~\cite{2019ApJ...877..116W}, we obtain
\be
N_{\rm H} = 3.63 \times 10^{21}\,{\rm cm}^{-2} \;  A_{\rm G}/{\rm mag} . 
\ee
In addition, we account for helium which contributes $9.1\%$ to the
number density of the ISM.

In order to check the completeness of this map, we calculate the
resulting average surface density $\Sigma$. Summing over the $z$ coordinate,
and multiplying by a factor $1.4$ to account for helium and heavier elements,
we obtain $\Sigma=10.4\,M_\odot/$pc$^2$. The comparison with the estimate
$\Sigma=13M_\odot/$pc$^2$ for the local surface density from
Ref.~\cite{Flynn:2006tm} indicates that the map includes $\simeq 80\%$
of the total gas.

\begin{figure}[tb]
    \centering
      \includegraphics[width=\columnwidth]{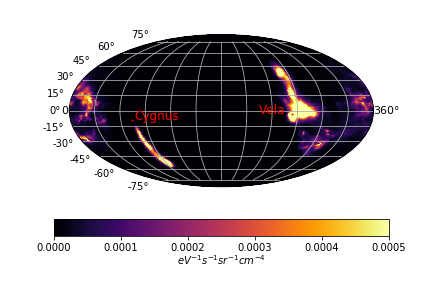}
    \caption{The function $\Xi(E,l,b)$  in Galactic coordinates
    for the proton energy $E=3$\,PeV. }
    \label{fig:Sigma}
\end{figure}

Combining the obtained matter map with the CR densities and integrating
along the line-of-sight, we define the function
\begin{align}
 \Xi^{A,A'}(E,l,b) = \int_0^\infty \!\!\d s\:  n^{A'}_{\rm gas}(\vec x) 
  I^{A}_{\rm CR}(E,\vec x) 
\end{align}
with $I^{A}_{\rm CR}(E)$ as the CR intensity of nuclei with mass number $A$.
The intensity of nuclei is in turn connected to their differential number
density as $I^A_{\rm CR}(E)=c/(4\pi) n^A_{\rm CR}(E)$. We find  in $\Xi^{pp}$
the two prominent hot spots visible in Fig.~\ref{fig:Sigma}: An
  extended region produced by CRs from the Vela SNR interacting with gas in the
  wall of the Local Bubble and another, smaller one produced by CRs from
  the SNR~G074.0-08.5 in the Cygnus Loop. The latter one is close in
  position with the  hottest spot in the IceCube neutrino data, and 
  we will therefore discuss this possible excess first.

\begin{figure*}[bt]
    \centering
      \includegraphics[width=\columnwidth]{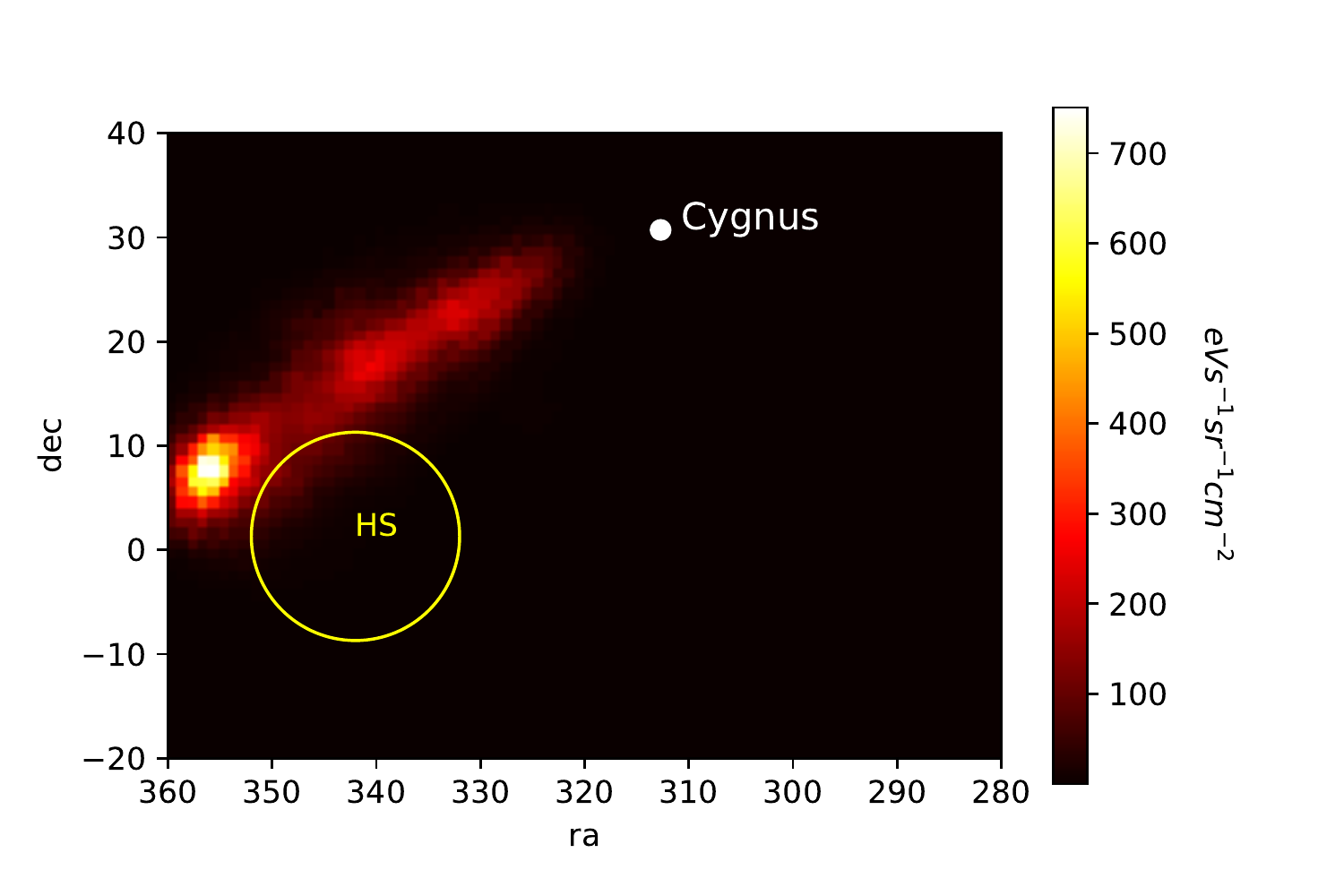}
       \includegraphics[width=\columnwidth]{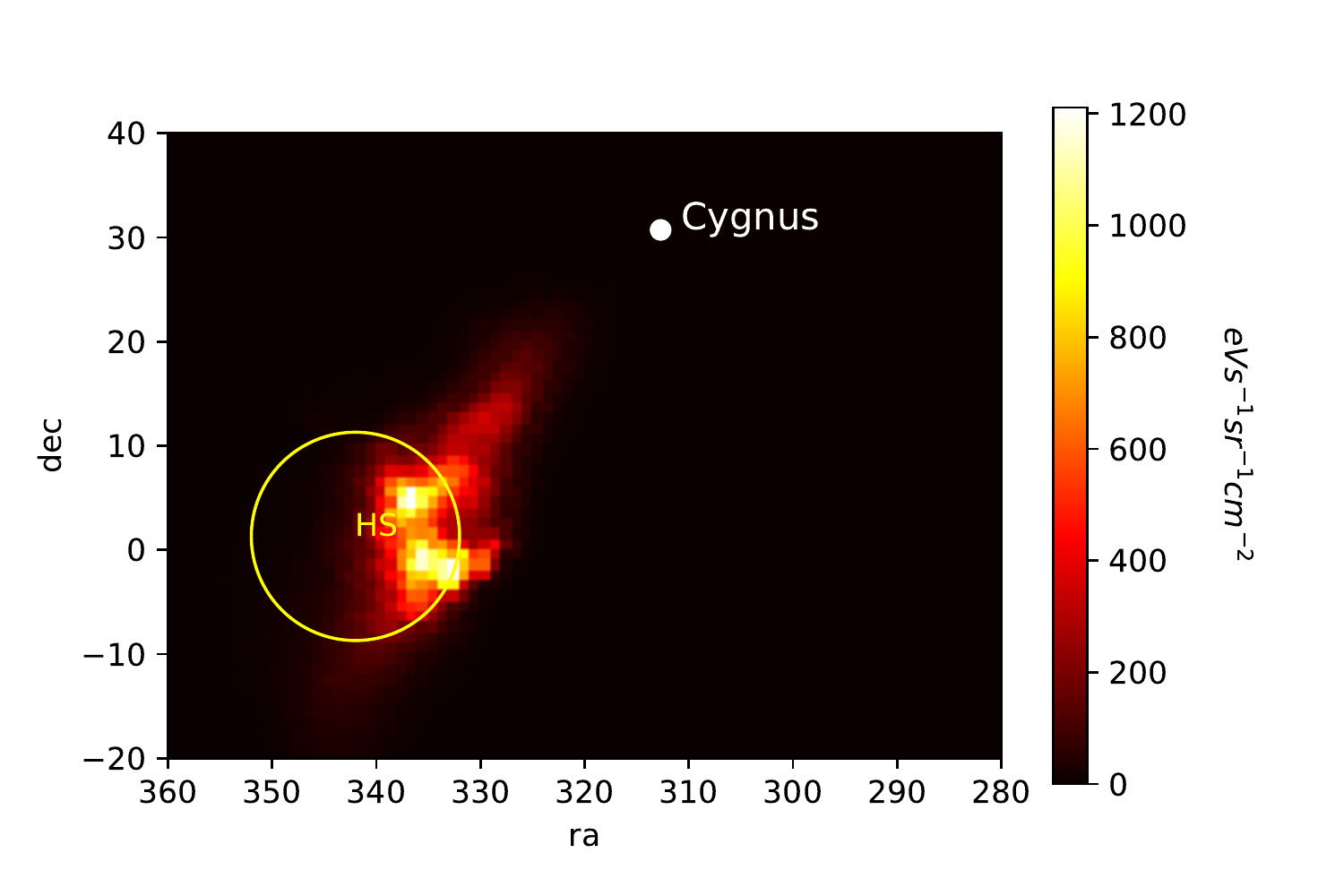}
       \caption{Neutrino intensity $E^2I_\nu(E,\alpha,\delta)$ in equatorial coordinates close to the hot
         spot for neutrino energies $E=100$\,TeV ;
         in the Janson-Farrar model
         (left) and the Prouza-Smida model (right).
         The yellow circle shows the most
         significant region in the IceCube search for point sources.
    \label{Ieq}}
\end{figure*}

\begin{figure}
\centering
  \centering
  \includegraphics[width=1.\linewidth]{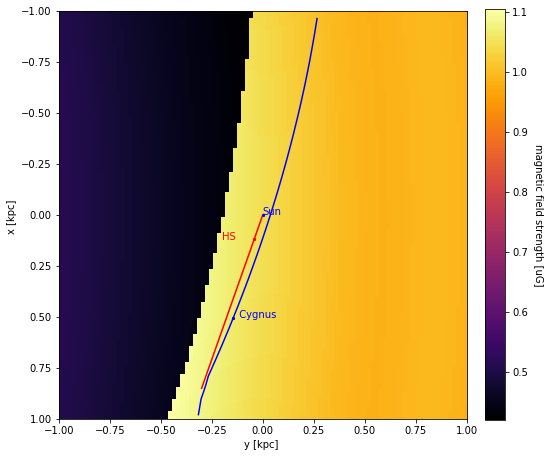}
  \includegraphics[width=1.\linewidth]{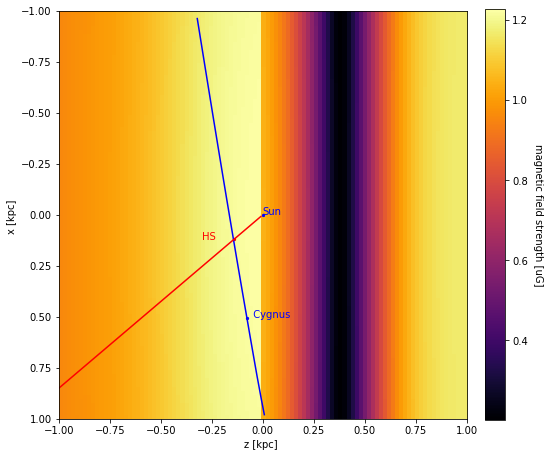}
\caption{The projection of the magnetic field line passing through the Cygnus loop SNR  is shown in blue and the position of the Sun in black. The red line represent the position of the IceCube hot spot for varying $R$, the red point corresponds to the position of the IceCube hot spot for $R_{hs} = 194$\,pc.}
\label{fig:MFcyn}
\end{figure}

As shown in Fig.~1 of the Appendix, CRs emitted by the Cygnus SNR
propagate along a magnetic field line 
away from the Galactic plane until they reach a region of increased
gas density close to the boundary of the Local Bubble where the interaction
probability $\propto \Xi(E,l,b)$ has a maximum. 
The position of the hot spot is approximately determined by
the intersection of the ellipsoid filled by CRs and the boundary of the Local
Bubble. From the left panel of Fig.~\ref{Ieq}, we see that the hot spot
predicted by us using the Jansson-Farrar model for the GMF is around
$10^{\circ}$--$15^{\circ}$ offset against the most significant point in the
search for point sources found by IceCube. However, rather small changes in
the direction of the magnetic field line connected to the Cygnus SNR could
reconcile the predicted and the observed position of the hot spot: 
In Fig.~\ref{fig:MFcyn}, we show the projection of the JF magnetic field
line passing through Cygnus (blue line) on the Galactic plane and perpendicular to the Galactic plane, together with the position of
the IceCube hot spot varying its distance as a red line.
  (Since the hot spot is observed on 2D, its 3D projection result in a line.)  
From the bottom panel of Fig.~\ref{fig:MFcyn}, we can see that the MF line passing through Cygnus crosses the hot spot line in the $x$-$z$ plane.
From the top panel of Fig.~\ref{fig:MFcyn}, it is
clear that a rather small rotation (anti-clockwise) by $\simeq 5^{\circ}$
of the magnetic field line in the Galactic plane  would be sufficient
to align CRs from the Cygnus loop SNR with the position of
the Ice Cube hot spot. 
Note also that small changes in the position of the hot spot would not lead
to large changes in the resulting neutrino intensity, because the gas density
in the hot spot is not atypically large, ${n_{\rm gas}}\simeq 0.6$/cm$^3$,
compared to other, close parts of the boundary of the Local Bubble.

In order to support this statement, we have also studied the resulting
neutrino hot spot in another GMF model, the one of Prouza and
Smida~\cite{Prouza:2003yf},
where we ajusted the ratio between the turbulent to regular magnetic field
in order to obey to B/C measurments.
The resulting neutrino intensity is shown in the right panel of
Fig.~\ref{Ieq}. One can see that the predicted hot spot in this model
overlaps nicely with the observed excess in the IceCube search for
point sources.

\begin{figure}[ht]
    \centering
      \includegraphics[width=\columnwidth]{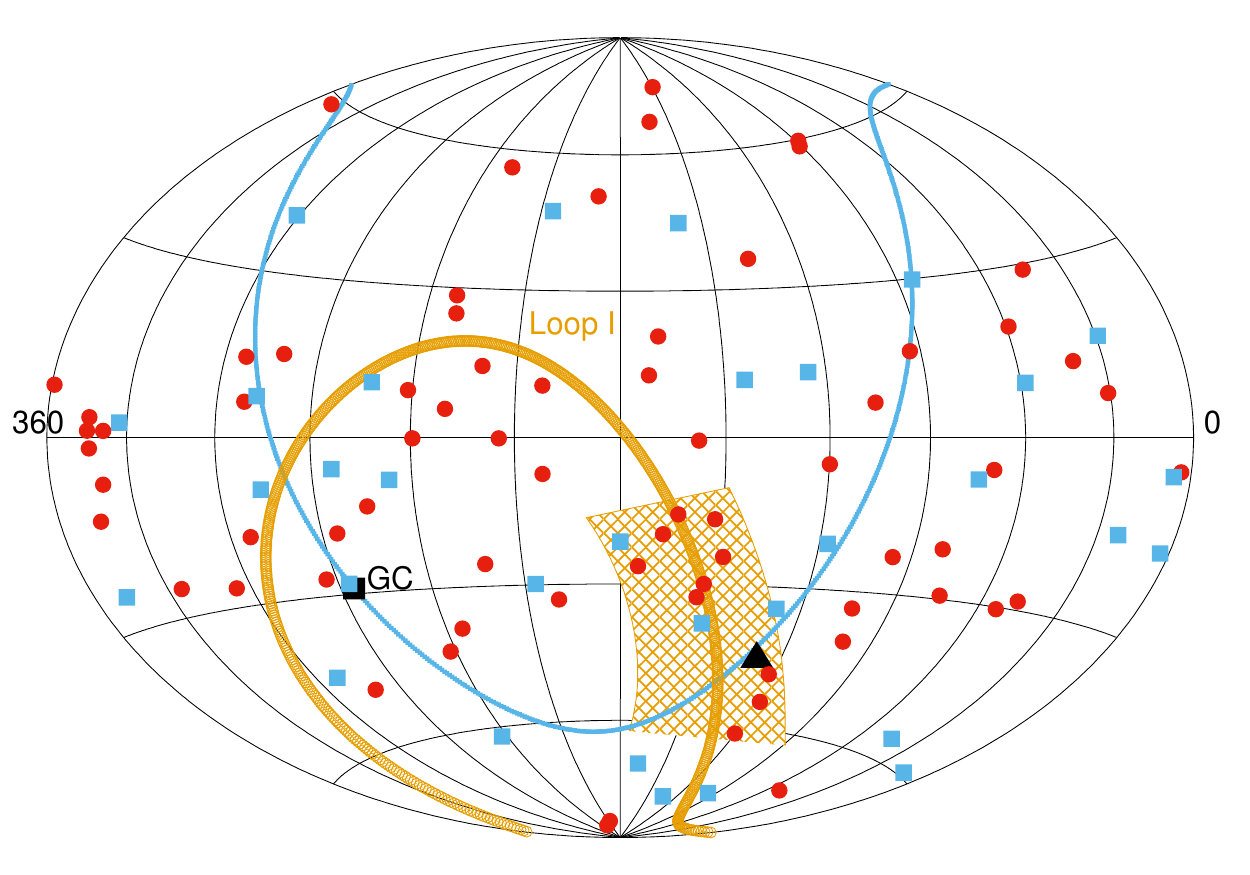}
      \caption{HESE (red dots) and muon (blue boxes) neutrino events,
        Vela (black triangle),  Loop~I
        and the chosen segment (hatched area) are shown
        in equatorial coordinates.}
    \label{fig:Loop}
\end{figure}

Finally, we want to discuss the second hot spot produced by CRs from Vela
which is predicted by our model. Although more prominent,
the corresponding neutrino signal is, at first sight,  absent in the
IceCube HESE sample. In order to investigate this issue more, we show
in Fig.~\ref{fig:Loop} the arrival directions of the public IceCube
neutrino events together with the position of Vela (black triangle) and
the superbubble Loop~I (orange circle). This superbubble interfaces with
the Local Bubble, while the Vela SNR is close to its boundary. Both bubbles
should lead to large deviations from the global regular field predicted 
in GMF models like the ones of Jansson-Farrar and Prouza-Smida. However,
the two GMF models used by us have been only modified to account for the
presence of the Local Bubble, while our calculations of CR trajectories do
not take into account the
effect of Loop~I. If CRs would be deflected by the enhanced magnetic field
in the wall of Loop~I similarly as in the wall of the Local Bubble, an
increase of neutrino events along the shell of Loop~I close to Vela should
result. As an illustration of this effect, we show in Fig.~\ref{fig:Loop}
the segment of Loop~1 close to the intersection of the magnetic field line
in the Jansson-Farrar model as a hatched area.
The number of HESE neutrino events in this area is ten, while six neutrinos
are expected. Thus one may speculate that this excess is connected to
CRs from the Vela SNR.

\section{Neutrino intensity}            

We concentrate first on the neutrinos produced by CRs from SNR~G074.0-08.5
in the Cygnus Loop. In this section, we report the results of our numerical
calculations
which we compare in the appendix with analytical estimates. Because of the
rather large uncertainty in the distance to this SNR,  $d=(0.5-1)$\,kpc,
we consider two cases, choosing as the distance the average ($d=0.75$\,kpc)
and the minimal value ($d=0.5$\,kpc) of this range, respectively, 
  and choosing for the GMF the JF model. The intensity
$I_\nu(E,l,b)$ of neutrinos with energy $E$  emitted along the
line-of-sight with direction $(l,b)$ is given by
\begin{align}
  I_\nu (E,l,b)= \sum_{A,A'}  \int_E^\infty \!\!\d E' \:
 \Xi^{A,A'}(E',l,b) \frac{\d\sigma^{AA'\to\nu}(E',E)}{\d E}.
\end{align}
 We include both for the target
gas and the CRs the contribution of protons and helium nuclei,
$A,A'=\{p,{\rm He}\}$. The neutrino production cross sections are evaluated 
with {\tt AAfrag}~\cite{Kachelriess:2019ifk}. In the left panel of
Fig.~\ref{Ieq}, we
show the resulting  neutrino intensity $I_\nu$ for the energy $E=100$\,TeV
in equatorial coordinates, i.e.\ as function of right ascension $\alpha$ and
declination $\delta$ for $d=0.5$\,kpc. For the larger
source distance,  $d=0.75$\,kpc, the hot spot in the neutrino intensity
shrinks slightly in size, while its position is nearly unchanged.
The maximum of $I_\nu(E,\alpha,\delta)$ corresponds to $\simeq 400$ times
the isotropic neutrino intensity measured by IceCube.
In the case of the larger source distance,  $d=0.75$\,kpc, the neutrino
intensity is reduced by one third.

\begin{figure}[t]
  \centering
     \includegraphics[width=\columnwidth]{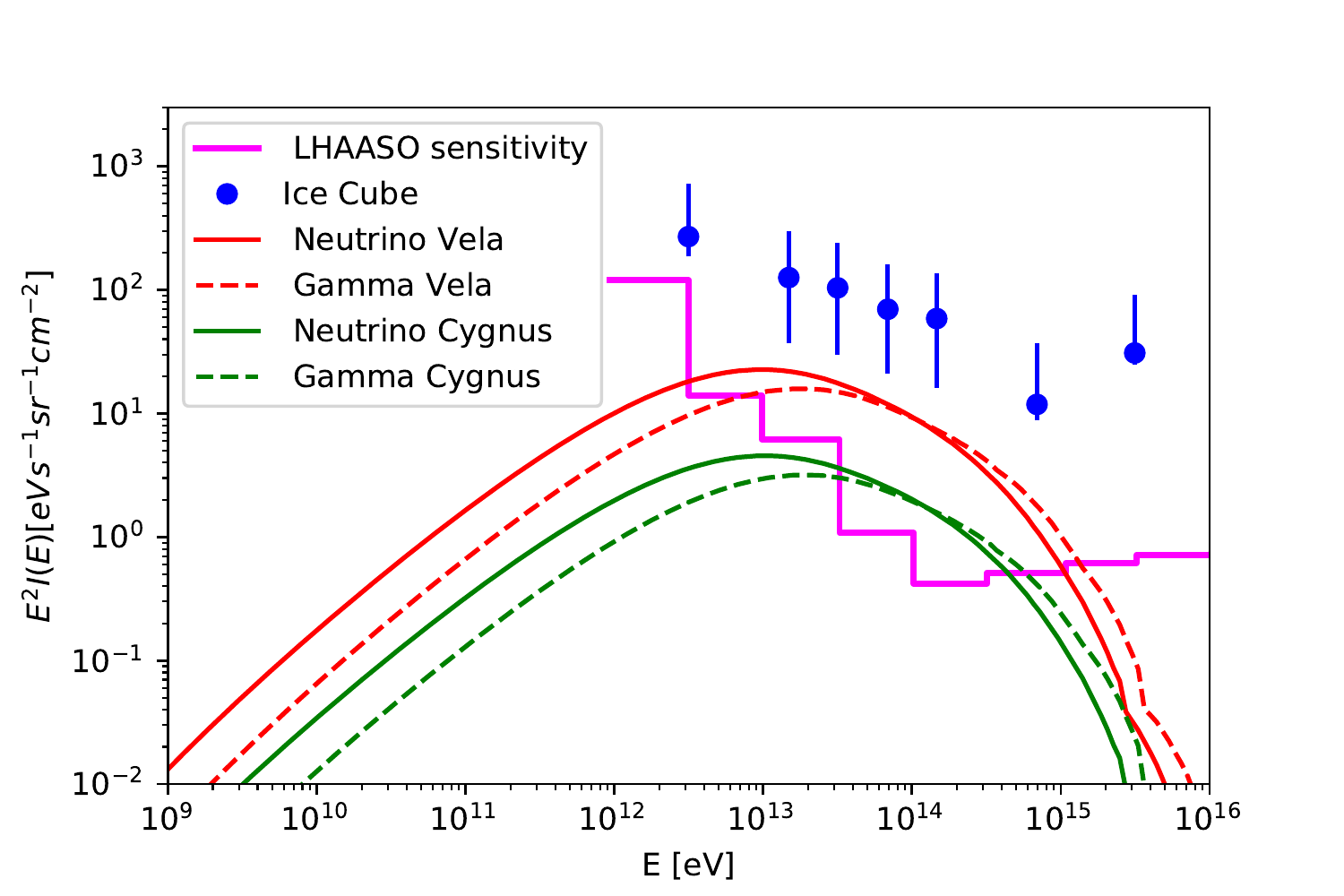}
    \caption{Equivalent isotropic intensity of a
      neutrinos and photons produced by CRs from the Cygnus Loop in green and Vela in red are compared to IceCube data~\cite{Aartsen:2017ujz} and the estimated sensitivity
      of the LHAASO experiment~\cite{Neronov:2020wir}.
    \label{fig:Iiso}}
\end{figure}

We define the equivalent isotropic intensity of a specific source as
\be \label{Isio}
\vev{I_\nu(E)} \equiv \frac{1}{4\pi} \int\d\Omega \, w(E) I_\nu(E,l,b) 
\ee
with the weight $w(E)$ set to one.
Since the observed neutrino intensity $I^{\rm obs}_\nu(E)$ is approximately
isotropic, the ratio $I_\nu(E)/I^{\rm obs}_\nu(E)$ corresponds to the
fraction of neutrino events with energy $E$ from a given source
observed by an experiment with  uniform exposure. In Fig.~\ref{fig:Iiso},
we show the isotropic neutrino intensity $\vev{I_\nu(E)}$ produced by CRs from
the Cygnus Loop together with Ice Cube data from Ref.~\cite{Aartsen:2017ujz}.
Note that the neutrino flux is
dominated by the contribution from helium primaries which have a larger
interaction probability and dominate above $E\simeq 10^{14}$\,eV the CR
injection spectrum defined in Eq.~(\ref{spec}). Additionally, we compare
the intensity of photons
to the sensitivity of the LHAASO experiment estimated in
Ref.~\cite{Neronov:2020wir}.

For a specific experiment like IceCube, we have to account for the
declination dependence of the effective area $A_{\rm eff}(E,\delta)$.
For the HESE data set, we use the the effective area $A_{\rm eff}(E)$
from Ref.~\cite{Abbasi:2020jmh} and deduce the declination
dependence
from  Ref.~\cite{Aartsen:2014gkd}, see App.~\ref{exp} for details.
Since the extension of the hot spot is small, we can neglect  the
declination dependence of the  weight, setting
$w=A_{\rm eff}(E,\delta_s)/A_{\rm eff}(E)\simeq 1 $
with $\delta\simeq 0^\circ$ for the hot spot of  IceCube.

We estimate the number of expected neutrino events above the minimal
energy $E_0$ as
\be \label{Nev}
N_\nu(E>E_0)  =  wT \int_{E_0}^\infty dE \int\d\Omega  A_{\rm eff}(E) I_\nu(E,l,b), 
\ee
where $T$ is the observation time. With $E_0=60$\,TeV, we find that 
the hot spot produced by CRs from the Cygnus loop corresponds to two neutrino
events, while four neutrino events in the HESE (event ID 44, 67, 74  with
corresponding energies of 84.6 \,TeV, 165 \,TeV, 71.3 \,TeV respectively, and
event 105) may be associated with hot spot found in the IceCube analysis.
Repeating these calculations for the GMF model of Prouza and Smida, the
estimated number of neutrino event increases to $\sim 3$--4.

In Fig.~\ref{fig:Iiso}, we show additionally the isotropic neutrino intensity
$\vev{I_\nu(E)}$ produced by CRs from Vela, using our default parameters
for the injected CR power and spectra. This intensity is around a factor of
six higher than the one from the Cygnus loop using the JF model and
would corresponds thus to 8--16 neutrino events. Hence, the CR power injected
by Vela would have to be reduced by a factor 2--3, or the CRs should be spread
over a larger segment of Loop~I, to be consistent with the IceCube
observations.

\section{Discussion}
\label{disc}

First, we should comment on the fact that our model contains various
uncertain parameters, leading to a corresponding spread in our predicitions
for the resulting secondary fluxes. We previously presented in
Ref.~\cite{Bouyahiaoui:2020rkf} the dependance of our results varying the
parameters of our magnetic field model for the Local Bubble. 
One should also take into account the uncertainties of the GMF model adopted:
In this case, we employed two differents models that gave similar results.
Additional uncertainties are connected to the dust maps,  that are of the
order of 10\%. Finally, variations in the injection spectrum, the composition
and the power of individual sources may affect our results for the secondary
fluxes.

In the previous sections, we have considered only SNRs with a distance
from the Sun smaller than 1\,kpc. Let us now verify that other SNRs similar
to the Cygnus Loop but at larger distances contribute a smaller number of
events to the IceCube data.  There are about ten such sources in a 1\,kpc
region around the Sun, which corresponds to  about 7000~sources in the Galaxy.
The Cygnus Loop SNR contributes about 2--4 events to the IceCube data set
due to the interactions of CRs with the Local Bubble at the distance of
$\simeq 100$\,pc. Thus the same source at 1\,kpc contributes
0.02--0.04\,events, bringing the contribution of nearby sources which are
not connected by a magnetic field line to the Sun to 0.2--0.4 events.
For sources at 10\,kpc distance, the contribution of a single source is
reduced to 0.0002--0.0004 events. With about 3000 of such sources, they
add 0.6--1.2 events. All together, we expect about one or two events in
IceCube from similar sources in the whole Galaxy, which is consistent with
the limits on the contribution from the Galactic plane deduced from the
IceCube data.

Next, we comment on the prospects for the  detection of gamma-rays
which are produced in association with neutrinos in the hot spot. 
Since the predicted gamma-ray emission from the hot spot is not point-like
but rather extended, its detection is challenging and depends strongly on
the rejection capability of hadron in the considered gamma-ray experiment.
HAWC can detect a diffuse gamma-ray flux on a level comparable to the
over-all diffuse neutrino flux measured by IceCube shown, but is not
sensitive enough to detect the expected  gamma-ray flux from the hot spot
shown in Fig.~\ref{fig:Iiso}. 
Limits on the gamma-ray flux at energies $E\sim 100$\,TeV can also
be derived from data of the ASgamma experiment~\cite{Amenomori:2021gmk}.
For instance,  in Ref.~\cite{Neronov:2021ezg} a limit on the equivalent
isotropic gamma-ray intensity was derived which can be compared to
flux from the hot spot presented in Fig.~\ref{fig:Iiso}. At 400\,TeV, the
expected flux is around $1\,{\rm eV/cm^2/s/sr}$, while the limit from
ASgamma is an order of magnitude higher. At 1\,PeV, the expected gamma-ray
flux is $0.3\,{\rm eV/cm^2/s/sr}$, while the limit from ASgamma is five times
higher. In contrast, the more sensitive LHAASO
experiment~\cite{Addazi:2019tzi}
should be able to detect the photon flux from the hot spot in the energy
range  $E\sim 100$\,TeV, but the required time of data taking
will depend on the extension of the hot spot: In the case of the JF model
shown in the left panel of Fig.~1, we can estimate the gamma flux
at 100\,TeV as
$E^2 {\cal F}_\gamma\simeq E^2I_\gamma\times\Omega\simeq 7\times 10^{-12}$erg/cm$^2$/s,
assuming a radius of $3^\circ$ of the hot spot with an intensity of $I_\gamma\simeq
500$eV/cm$^2$/s/sr. Rescaling the sensitivity $S$ of LHAASO for point
sources, $S_{\rm point}$, as $S=S_{\rm point}\delta/\delta_{\rm PSF}$,
where $\delta_{\rm PSF}\simeq 0.3^\circ$ is the extension of the
point-spread function and $\delta$ the one of the extended source,
it follows $S\simeq 4\times 10^{-13}$erg/cm$^2$/s. Although being
very crude, this estimate suggests that such a concentrated hot spot
should be easily detectable for LHAASO within one year of data taking.
The hot spot is more difficult to detect if its extension, for the
same flux, is wider, since then the CR background increases. Such
a larger extension is expected in the PS model, or more generally,
for a relative increase of the turbulent to the regular magnetic
field. Moreover, the larger extension would be still consistent
with the one of the IceCube's hot spot.

One should not confuse the Cygnus superbubble which is located in Galactic
plane~\cite{Abeysekara:2021yum,Dzhappuev:2021cfj}  with the Cygnus Loop SNR,
considered here. In the case of sources located in the Cygnus superbubble,
one expects secondary gamma-rays and neutrinos in the same region. In
contrast, the neutrinos and gamma-ray emission considered here is
produced after CRs streaming away from the Galactic disk
hit gas clouds close to the boundary of the Local Bubble.

Finally, let us comment on the recent detection of 0.1--1\,PeV diffuse
gamma-rays reported by the Tibet ASgamma collaboration~\cite{Amenomori:2021gmk}.
In their event sample, all gamma rays with energies above 398\,TeV were
separated by more than $0.5^\circ$ from known TeV sources.  In the scenario
of anisotropic CR diffusion discussed here, such a separation arises rather
naturally. This measurement combined with Fermi LAT data at TeV energies
indicate also that the photon flux in the outer Galaxy is rather
hard, $1/E^{2.5}$~\cite{Koldobskiy:2021cxt}.

\section{Conclusions}
\label{concl}

We have proposed in this work that CRs from the SNR~G074.0-08.5 in the
Cygnus Loop interacting with gas close to the boundary of the Local
Bubble lead to an hot spot in the neutrino flux. Anisotropic
diffusion of CRs is a necessary ingredient of this proposals, leading
to enhanced CR densities along magnetic field lines connected to
CR sources. The position of this hot spot is compatible with the most
significant point in the IceCube search for point sources, and the
neutrino intensity estimated by us corresponds to two to four
events with energy above 60\,TeV.
In addition, we have proposed that CRs from the Vela SNR are
spread out by the magnetic field in the boundary of the superbubble Loop~I,
leading to an excess of neutrino events in a segment of Loop~I directed
towards Vela.
The corresponding photon fluxes should be detectable by LHAASO soon,
providing a clear signature for a Galactic origin of the hot spot
in the neutrino flux observed by IceCube.

\acknowledgments

We would like to thank Austin Schneider for comments on the IceCube exposure,
and Torsten En{\ss}lin and especially Raimer Leike for advice on the use of
their extinction maps. The work of D.S.\ was supported in part by 
the Ministry of Science and Higher Education of the Russian Federation
under the contract~075-15-2020-778 in the framework of the program 
``Large Scientific Projects'' within the national project ``Science''.



\appendix

\section{Comparison with analytical estimates}

\subsection{Cosmic ray density}

For our analytical estimates,            
we assume that the  Cygnus Loop SNR (G074.0-08.5) injected instantaneously
$6\times 10^{50}$\,erg in CRs, following a power law with
$Q(E)=Q_0 (E/E_0)^{-\alpha}$ and $\alpha\simeq 2.2$ for the injection spectrum.
To be definite, we split the total energy between protons and helium nuclei
as 4:3. Choosing the lowest injection energy of protons as the
normalization energy, $E_0=1$\,GeV,
it follows then $Q_p=E_p/(5E_0^2)\simeq 4.3\times 10^{52}/$GeV.
Similarly, it follows $Q_{\rm He}=4E_{\rm He}/(15E_0^2)\simeq Q_p$ for the
normalization of the helium flux above the minimal injection energy 4\,GeV.

The diffusion approximation can be applied once CRs  have reached distances
from the source that are greater than a few times the coherence length of
the turbulent magnetic field~\cite{Giacinti:2012ar,Giacinti:2013uya},
which is around $L_{\rm coh}\approx 10$\,pc close to the disk.
At a given energy,
the functional behavior of the observed CR flux from a single source
at the distance $L$ and with the age $\tau$ can be divided into three regimes: 
For $2D \tau\ll L^2$, the diffuse flux is exponentially suppressed, while for
intermediate times it is given by
\be \label{intens}
 I(E)\simeq \frac{c}{4\pi} \frac{Q(E)}{V(t)} .
\ee
Here,  $V(t)= 4\pi\, L_\perp^2 L_\|/3$ is the volume
of the ellipsoid with major axis $L_\|=(2D_\|t)^{1/2}$ and minor
axis  $L_\perp=(2D_\perp t)^{1/2}$, When the diffusion front reaches 
the edge of the Galactic CR halo, CRs start to escape and the slope of the
CR intensity steepens.

For the numerical values of the diffusion coefficients in the case of
anisotropic diffusion, we read from Fig.~4 from Ref.~\cite{Giacinti:2017dgt}
with $\eta=0.25$ and $D_{\rm iso}\simeq 1\times 10^{30}$cm$^2/$s valid at
the reference energy $E_\ast=10^{14}$\,eV that $D_\|\simeq 5 D_{\rm iso}$,
while $D_\perp\simeq D_{\rm iso}/500$. Hence CRs with energy $E_\ast$
fill an ellipsoid with major axis $L_\|\simeq 700$\,pc and minor
axis  $L_\perp\simeq 14$\,pc. The CR intensity of protons inside this ellipsoid
can be estimated with $V\simeq 1.7\times 10^{61}$cm$^3$ as
$E_\ast I(E_\ast )\simeq 6\times 10^{-6}/({\rm cm^2 \,s \,sr} )$.

\begin{figure}
    \centering
    \includegraphics[width=\columnwidth]{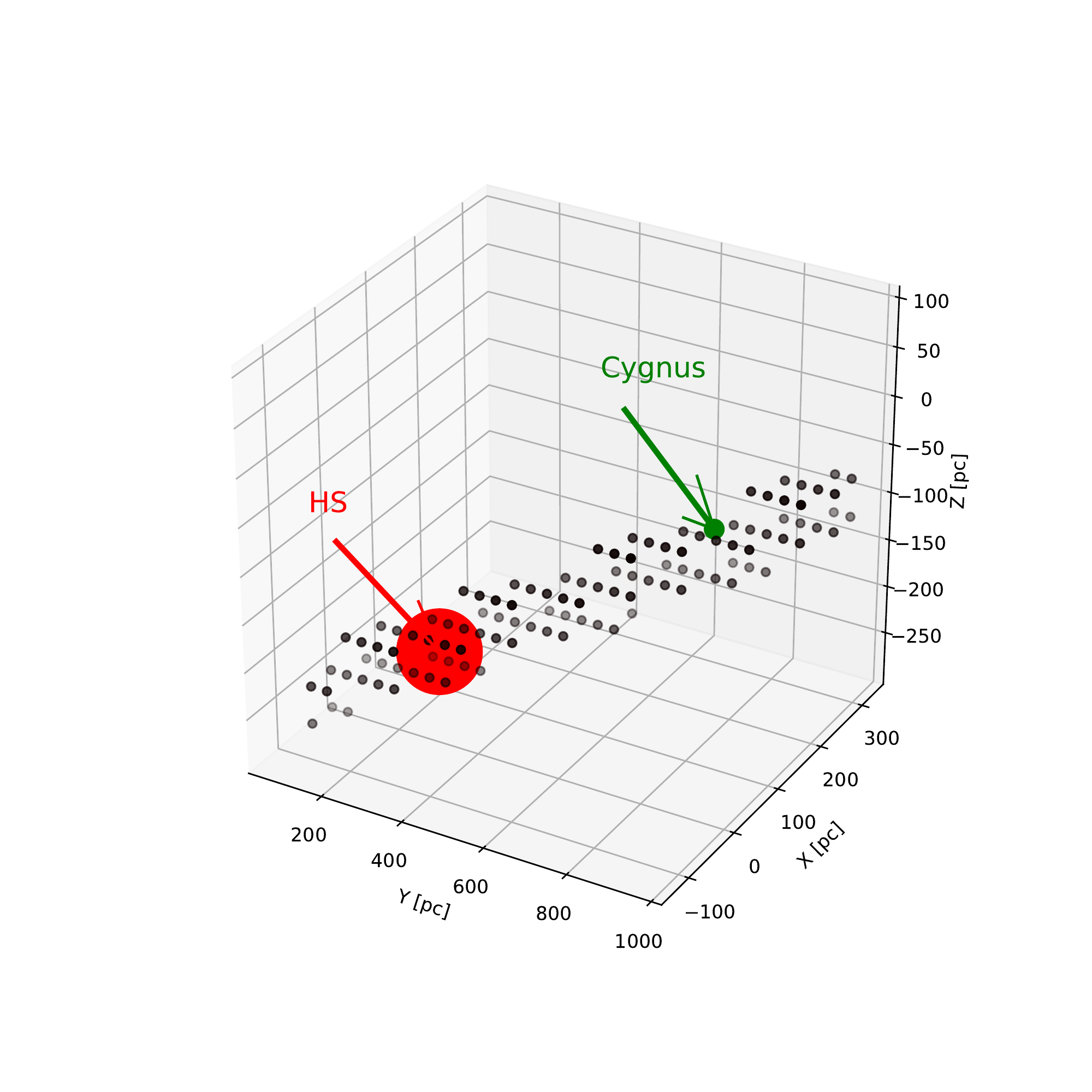}
    \caption{Volume satisfying $I>0.01I_{\max}$ at the proton energy
      $E=10^{14}$\,eV
      together with the position of the hot spot
      and the Cygnus Loop SNR.
    \label{tube}}
\end{figure}

In Fig.~\ref{tube}, we show for comparison the cells satisfying
the condition $I(E)>0.01I_{\max}(E)$ at $E=10^{14}$\,eV in our numerical
simulations together with the position of the hot spot and of the Cygnus
Loop SNR.
One can see that the CRs fill a tube with diameter 50\,pc, what agrees quite
well with the expectation $\sim 6\times 2D_\perp\tau\simeq 80$\,pc for
$I(E)=0.01I_{\max}(E)$. Moreover, the distance between the hot spot and the
Cygnus Loop SNR is around 530\,pc and the CR tube extends until it touches
the Local Bubble.
Note also this diameter is much larger than the extension of a 
SNR at the age of few hundred years; thus our assumption of a point-like
injection is justified.

\subsection{Neutrino production}            

In order to obtain an estimate for the  equivalent isotropic neutrino
intensity, we define
first the volume $\V$ of the neutrino emitting region by the condition
\be  \label{condition}
n_{\rm gas}(\vec x)  I_{\rm CR}(E_\ast,\vec x) >
0.01\max_{\vec x \in\text{box}}\{n_{\rm gas}(\vec x)  I_{\rm CR}(E_\ast,\vec x)\}
\ee
at a given energy $E_\ast$. Then we introduce in\footnote{Note that a factor
  $1/(4\pi)$ was erroneously missed in the corresponding formulas of
  Refs.~\cite{Andersen:2017yyg,Bouyahiaoui:2020rkf}.}
\begin{align}
 I_\nu(E) & = \frac{c}{4\pi} \sum_{A,A'\in\{1,4\}} \int_E^\infty{\rm d}E' \,
          \frac{{\rm d}\sigma_{\rm inel}^{AA'\to\nu}(E',E)}{{\rm d}E}
\\ & \times
   \int{\rm d}^3x \, \frac{n_A(E',\vec x) n_{\rm gas}^{A'}(\vec x)}{4\pi d^2}\,,
\nonumber     
\end{align}
the inelastic cross section $\sigma_{\rm inel}^{AA'}$, the spectrally averaged
energy fraction $\vev{y^{\alpha-1}}$ transferred  to neutrinos, and the total
number of gas particles 
$N_{\rm gas} = \sum_{A} \int_{\V} {\rm d}^3x \, n_{\rm gas}^{A}(\vec x)$ in
the source region $\V$. Moreover, we neglect the extension of the source
region, obtaining
\be
I_\nu(E)  \simeq \sum_{A,A'}  \frac{f_{A'} N_{\rm gas}}{4\pi d^2}\,
 \vev{y^{\alpha-1}} \sigma_{\rm inel}^{AA'\to\nu}(E)  \vev{I_A(E)} ,
\ee
where $d$ denotes the distance to the source volume $\V$,
$f_A$ the fraction of proton and helium nuclei in the gas,
and $\vev{I_A(E')}$ the spatially averaged intensity of CR protons and helium
nuclei.

At the energy $E_\ast=10^{14}$\,eV,  the condition~(\ref{condition}) is
satisfied in 750~cells of size (6\,pc)$^3$, resulting with
$\vev{n_{\rm gas}}\simeq 0.56$/cm$^3$ into $N_{\rm gas}\simeq 2\times 10^{60}$.
The intensity of CR protons  obtained from our numerical simulations and
averaged over this volume is
$\vev{E_\ast I_p(E_\ast)}\simeq 4\times 10^{-6}/$cm$^2$/s/sr,
i.e.\ agrees well with our estimate using the diffusion approximation in the
previous subsection.

With $d\simeq 270$\,pc and
$Z(E_\ast,\alpha)=\vev{y^{\alpha-1}}\sigma_{\rm inel}^{pp}\simeq 7$\,mbarn
for the $Z$ factor for the production of neutrinos from a power law
with $\alpha\simeq 2.2$, this leads to
$$
E_\ast^2 I^{pp}_\nu(E_\ast)\simeq 1.0 {\rm \frac{eV}{cm^2\,s\,sr}}
$$
for the isotropic neutrino intensity due to $pp$ interactions.
We can estimate the contribution of helium projectiles and targets
at energies $E_\nu\lsim 10^{13}$\,eV setting  $I_p(E)\simeq I_{\rm He}(E)$
valid for primary energies $\lsim 10^{14}$\,eV. With
$\sigma_{\rm inel}^{pp}\simeq 48$\,mbarn,
$\sigma_{\rm inel}^{p{\rm He}}\simeq 148$\,mbarn,
$\sigma_{\rm inel}^{{\rm He}p}\simeq 137$\,mbarn and
$\sigma_{\rm inel}^{{\rm He\,He}}\simeq 324$\,mbarn, it follows
then
\be
\frac{I_\nu^{\rm tot}}{I_\nu^{\rm pp}}\simeq
\frac{ 0.9(\sigma_{\rm inel}^{pp}+\sigma_{\rm inel}^{{\rm He}p} )+
       0.1(\sigma_{\rm inel}^{{\rm He}p} + \sigma_{\rm inel}^{{\rm He\,He}}) }
     {\sigma_{pp}}\simeq 4 .
\ee
Thus our estimate agrees again well the numerical value shown in
Fig.~\ref{fig:Iiso}.

\section{Exposure}  \label{exp}

We subtract in the Supplemental Fig.~5 of Ref.~\cite{Abbasi:2020jmh} the line
signal+background from the background to obtain the signal $S(\delta)$ as
function of the declination $\delta$. Then we calculate with $x=\sin\delta$
\be
 S_{\rm av}= \frac{1}{2} \int_{-1}^1 {\rm d}x \,S(\delta) .
\ee
The weight $w$ of a source with declination $\delta$ follows then
as  $w=S(\delta)/S_{\rm av}$.


\end{document}